\documentclass[
    5p, 
]{elsarticle}

\biboptions{sort&compress}
\usepackage{graphicx}
\usepackage{amsmath}
\usepackage{amssymb}

\usepackage{xcolor}

\newcommand{\MS}{\reflectbox{S}}

\journal{Extreme Mechanics Letters}

\begin{document}

\begin{frontmatter}
\title{Accelerated snapping of slender beams under lateral forcing}

\author[lion,amolf]{Colin M. Meulblok\corref{cor}}
\ead{meulblok@physics.leidenuniv.nl}

\cortext[cor]{Corresponding author}

\author[amolf,LAUM]{Hadrien Bense}

\author[amolf,ENS]{M. Caelen}

\author[lion,amolf]{Martin van Hecke}

\affiliation[lion]{
    organization={Huygens-Kamerlingh Onnes Laboratory, Universiteit Leiden},
    addressline={PO Box 9504},
    city={Leiden},
    postcode={2300 RA},
    country={the Netherlands}
}
\affiliation[amolf]{
    organization={AMOLF},
    addressline={Science Park 104},
    city={Amsterdam},
    postcode={1098 XG},
    country={the Netherlands}
}

\affiliation[LAUM]{
    organization={Laboratoire d’Acoustique de l’Université du Mans (LAUM),
UMR 6613, Institut d’Acoustique - Graduate School (IA-GS),
CNRS},
    addressline={Le Mans Université},
    City={Le Mans},
    postcode={72085},
    country={France}
    }
\affiliation[ENS]{
    organization={Laboratoire de Physique de l’École Normale Supérieure, CNRS, PSL Research University, Sorbonne Université, Université de Paris},
    City={Paris},
    postcode={F-75005},
    country={France}
}

	
\begin{abstract}
The hysteretic snapping under lateral forcing
of a compressed, buckled beam  is fundamental for many devices and mechanical metamaterials. For a single-tip lateral pusher,
an important limitation is that
snapping requires the pusher to cross the centerline of the beam. 
Here, we show that dual-tip pushers allow {\em accelerated snapping}, where the beam snaps before the pusher reaches the centerline.
As a consequence, we show that when a buckled beam under increased compression comes in
contact with a dual-tip pusher, it
can snap to the opposite direction --- this
is impossible with a single-tip pusher.
Additionally, we reveal a novel {\em two-step snapping} regime, in which the beam sequentially loses contact with the two tips of the dual-tip pusher.
To characterize this class of snapping instabilities, we
employ a systematic modal expansion of the beam shape. 
This expansion allows us to
capture and analyze 
the transition from one-step to two-step snapping geometrically. Finally we demonstrate how to maximize the distance between the pusher and the beam's centerline at the moment of snapping.
Together, our work opens up a new avenue for quantitatively 
and qualitatively
controlling and modifying
the snapping of 
buckled beams, with potential applications in mechanical sensors, actuators, and metamaterials. 
\end{abstract}
	
\end{frontmatter}

\section{Introduction}

Snap-through instabilities are widely observed in nature. The Venus flytrap and carnivorous waterwheel plants capture prey through snapping leaves~\cite{WaterwheelPRE,VenusFlyNat,VenusFlySci}. Hummingbirds exploit a snap-through mechanism in their beak to catch insects mid-flight~\cite{HummingbirdJTB}. Grasshoppers and click beetles leverage similar mechanisms to power their leaps~\cite{LocusJumpingJEB,LucusJumpingJIB,WangPNAS2023}.
Similar snap-through instabilities have been harnessed in mechanical metamaterials to realize a wide range of functionalities~\cite{BertoldiReview2017,HuReview15}, 
including microfluidic passive valves~\cite{GomezRPL2017}, microlens 
shells~\cite{HolmesAdvMat2007}, soft actuators~\cite{OverveldePNAS2015}, components of
soft robots~\cite{ZhangSoftRob2022,LishuaiAdvIntSyst2023}
and counter-snapping ~\cite{Ducarme2025}.
Often, snap-through is triggered by a lateral pusher that forces the beam into an S-shaped configuration at the onset of snapping~\cite{Chen21,Vangbo98,Pandey14}. 
As a consequence, the lateral pusher has to cross the beam's centerline to induce snapping. This inherent geometric constraint limits design flexibility and hinders potential applications in fields such as {\em in-material} computing~\cite{Whitesides19,Yang2016,Yang2019,Ding2022,Kwakernaak2023,BensePNAS21,LiuPNAS24}.

Here, we show that introducing a second pusher tip provides the desired control and enhances design flexibility. Dual-tip pushers enable {\em accelerated snapping}, in which snapping does not require the 
pusher to cross  the centerline --- and this allows
snapping under compressive loading in the presence of a fixed, dual-tip pusher (Sect.~\ref{sec:pheno}). 
For such pushers, we uncover a novel {\em two-step snap-through} mechanism, in which the beam passes through a stable intermediate state before fully snapping (Sect.~\ref{sec:pheno}).
To elucidate the emergence of these behaviors, 
we develop a reduced-order numerical  model based on a modal expansion of the beam shape (\cite{Vangbo98,Pandey14,LazarusBook}, Sect.~\ref{sec:model}). 
We use this model to visualize the deformation energy of the beam
in a two-dimensional landscape, where lateral pushers
translate to excluded zones, and where beam configurations
in contact with the pusher reside precisely on the boundary of these zones. This allows a geometric interpretation of the stability and snap-through behavior of the beam. In particular, it reveals
that the second pusher tip introduces additional local extrema in the energy landscape, which underlie both the accelerated and two-step snapping behavior (Sect.~\ref{sec:results}). 
Finally, we systematically map the snapping threshold as a function of the geometry of the dual-tip pusher, allowing precise control over the onset and nature of the instability (Sect.~\ref{sec:results}).
Together, our work uncovers and characterizes a new class of snap-through behavior with potential applications in soft robots, smart sensors and actuators, metamaterials and {\em in materia} computing.

\begin{figure}[t]
    \centering
    \includegraphics[]{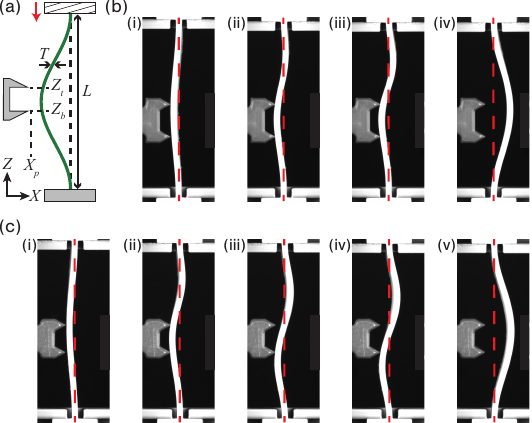}
    \caption{\textbf{Phenomonology} 
    (a) Schematic of a slender beam with rest length $L_0$, compressed to end-to-end distance $L$. The dual-tip pusher is defined by the vertical location of its tips, $Z_t$ and $Z_b$, and their horizontal offset from the beam centerline, $X_p$.
    (b-c) Experimentally observed snap-through transitions induced by increasing compression using fixed dual-tip pushers     
    and two values of $Z_b$.
    ($L_0=100$ mm, $T=3$ mm, $X_p=-5$ mm, $Z_t=2$ mm; see SI for details).
(b) One-step snapping at  for $Z_b=-16$ mm, (i-iv: $\varepsilon=0.011$, $0.022$, $0.027$, and $0.028$).
    (c) Two-step snapping for $Z_b=-13$ mm: here the beam loses contact with the tips sequentially  (i-v: $\varepsilon=0.011$, $0.025$, $0.026$, $0.028$, and $0.029$).
    }\label{fig:1}
\end{figure}

\section{Phenomenology} \label{sec:pheno}
We consider a slender beam 
under quasistatic compression
in the presence of a dual-tip lateral pusher. The beams have rest length \( L_0 \) with rectangular cross-sections (in-plane thickness $T$, out-of-plane thickness $W$) in the slender beam limit $T\ll W \ll L_0$. 
Compression is applied by
controlling the end-to-end distance $L$, defining the axial strain $\varepsilon=(L_0-L)/L_0$.
The pusher consists of two vertically spaced tips located at vertical coordinates $Z_t$ and $Z_b$, both sharing a common horizontal coordinate $X_b$ (Fig.~\ref{fig:1}a).

We begin by examining the qualitative evolution of a buckled beam as it makes contact with a fixed dual-tip pusher, rigidly connected to the bottom plate, under increasing compression (Fig.~\ref{fig:1}).
The beam is initialized in its left buckled state without contact with the pusher tips.
As $\varepsilon$ increases, the beam and pusher tips make contact, resulting in higher-order deformation modes. 
For $Z_b=-16$ mm, the beam deforms smoothly toward a critical configuration (Fig.~\ref{fig:1}bi-iii), and then experiences a single, abrupt snap-through instability (Fig.~\ref{fig:1}biii-iv). 
For $Z_b=-13$ mm, the snap-through occurs in two-steps: first
the beam loses contact with one tip and transitions
to an intermediate state tip (Fig~\ref{fig:1}ci-iii), and then
it loses contact with both tips and snaps to the right-buckled configuration (Fig~\ref{fig:1}civ-v).

We note that the emergence of snapping without the pusher crossing the beam’s centerline is linked to the moment applied by the pair of tips which forces the beam into a configuration unattainable with a single-tip pusher. A precise theoretical understanding of how these snap-through scenarios arise is the focus of the remainder of this paper.

\section{Theoretical modeling}\label{sec:model}
Our strategy to model snapping induced by 
dual-tip pushers
starts from a standard expansion of the beam shape.
First, we observe and explain why the first three modes in this expansion are dominant and focus on these. 
Then, using the constraint of fixed beam-length, we map our beams to 
a two-dimensional energy landscape. 
The lateral constraints translate to excluded zones, and beam configurations in contact with the lateral pusher exactly lie on the boundary of these zones.
As we show below, this facilitates a geometric interpretation of both one-step and two-step snapping, offering a clear physical understanding of these phenomena and their dependence of the pusher parameters.

In this section, we first derive the modal expansion (Sec.~\ref{sec:modal}). 
We then introduce a two-dimensional visualization of the energy landscape
(Sec.~\ref{sec:others}). 
We end this section by presenting a geometric picture for the evolution of the beam configuration, and illustrate this for
a classical singe-tip pusher (Sec.~\ref{sec:evol}).

\subsection{Modal expansion and beam model} \label{sec:modal}
We derive the modal expansion for the beam shape. 
We parametrize this shape with coordinates $X(Z)$, and
compress the beam via boundary conditions $X(\pm L/2) = 0$ and $ X'(\pm L/2) = 0$.
We focus on small deformations,
neglect friction between beam and tips, and consider the beam as inextensible. 
The bending energy per unit width then reads:
\begin{equation}
\mathcal{E}_b = \frac{B}{2}\int_{-L/2}^{L/2}C(Z)^2dZ~,\label{eq:Eb}
\end{equation}
with bending modulus $B = {Et^3}/{12(1-\nu^2)}$, Young's modulus $E$, Poisson ratio $\nu$ and curvature of the beam $C = {X''(Z)}/{(1+X'(Z))^{3/2}} $.
Under the small strain approximation, the curvature can be written as $C(Z) \approx X''(Z)$.

We now impose two sets of constraints on this problem. First, 
inextensibility imposes that the length of the beam is conserved:
\begin{equation}
\int_{-L/2}^{L/2}\sqrt{1+X'(Z)^2}dZ = L_0~. \label{eq:Lconst}
\end{equation}
Under the small strain approximation, Eq.~(\ref{eq:Lconst}) can be expanded to yield:
\begin{equation}
L + \frac{1}{2}\int_{-L/2}^{L/2}X'(Z)^2dZ = L_0. \label{eq:Lconst2}
\end{equation}
Second, the pusher tips constrains the local maximum deflection of the beam:
\begin{equation}
\begin{cases}
X(Z_t) < X_p~,\\
X(Z_b) < X_p.\label{eq:Hconst}
\end{cases}
\end{equation}

Combining the energy and constraints, we obtain 
the Lagrangian of this system:
\begin{equation}
\begin{split}
\mathcal{L}[X] = \frac{B}{2}\int_{-L/2}^{L/2}X''(Z)^2dZ\\
- P\left( L-L_0+\frac{1}{2}\int_{-L/2}^{L/2}X'(Z)^2dZ\right)\\
 - F_t\left(X_p - X(Z_t)\right) - F_b\left(X_p - X(Z_b)\right) \label{eq:Lagrange},
\end{split}
\end{equation}
where $P, F_t, F_b$ are the Lagrange multipliers and can be interpreted, respectively, as the axial force and top and bottom lateral forces per unit width. Importantly, we note that Eq.~(\ref{eq:Hconst}) are {\em inequalities}, and we deal with these using the Karush-Kuhn-Tucker conditions \cite{LazarusBook}.

We scale out the compressive strain \cite{Pandey14}, introducing the rescaled parameters:
\begin{equation}
\begin{cases}
z = Z/L = Z/(L_0(1-\epsilon)),\\
x = X/\left(L\sqrt{\epsilon}\right),\\
p = \frac{P}{B/L^2},\\
f = \frac{F}{B/L^2}\frac{1}{\sqrt{\epsilon}}\label{eq:rescale},
\end{cases}
\end{equation}
leading to:
\begin{equation}
\begin{split}
\mathcal{L}[x] = \frac{1}{2}\int_{-1/2}^{1/2}x''(z)^2dz\\
- p\left(\frac{1}{2}\int_{-1/2}^{1/2}x'(z)^2dz-1\right)\\
- f_t(x_p - x(z_t)) - f_b(x_p - x(z_b)) \label{eq:Lagrange_norm}.
\end{split}
\end{equation}
The advantage of this rescaling is that solutions for the buckled beam shape at different strains can be mapped onto each other using the rescaling. 
The price we pay is that the rescaled location of a fixed lateral pusher becomes strain dependent.

Finally, we perform an expansion of the beam shape on its first $m$ modes, i.e., write $x(z) \approx \sum_1^m a_i\xi_i(z)$, with ${\xi_i}$ the $i^{th}$ order solution of the linearized Elastica problem with clamped-clamped boundary conditions\cite{Vangbo98}:
\begin{equation}
\begin{cases}
\xi_i =- \frac{2}{n}(1 - \frac{\cos(nz)}{\cos(n/2)}),& \text{for }\sin(n/2)=0\\
\xi_i = -\frac{2}{n}(\frac{\sin(nz)}{\sin(n/2)} - 2z),& \text{for } \tan(n/2) = n/2,
\end{cases}\label{eq:modalenergy}
\end{equation}
where the values of $n$ are ordered in an increasing manner, and are assigned an index $i$ starting at $1$. This projection significantly simplifies Eq.~(\ref{eq:Lagrange_norm}):
\begin{equation}
\begin{split}
\mathcal{L}[a_1,a_i...a_m] = \sum_1^mn_i^2a_i^2 - p\left(\sum_1^ma_i^2 -1\right)\\
 -f_t\left(x_p-\sum_1^ma_i\xi_i(z_t)\right) - f_b\left(x_p-\sum_1^ma_i\xi_i(z_b)\right).
\end{split}\label{eq:Lagrange_final}
\end{equation}
The problem now consists in finding the set of coefficients $a_i$ that minimizes $\sum_0^m n_i^2a_i^2$, under the constraints $\sum_1^ma_i^2 -1=0$ and $\sum_1^m a_i\xi_i(z_k)-x_p\leq0$ (with $k=t,b$).

\subsection{Energy landscape} \label{sec:others}

To visualize the elastic energy landscape, 
we truncate the modal expansion to include only the dominant modes. 
The modes are ordered by their energy contribution (Eq.~\ref{eq:modalenergy}) and the minimal number of modes can be understood through a simple constraint-counting argument: the fixed length imposes one constraint, and each pusher tip adds another. Thus, a single-tip pusher requires at least two modes, while a dual-tip pusher requires at least {\em three} to satisfy the equations of motion. We confirm both numerically and experimentally that three modes are sufficient to capture the beam's evolution and that, when using more modes, the first three modes dominate the expansion (see SI).

We leverage the energy concentration in the first three modes, together with the inextensibility (length) constraint, to construct a strain-independent two-dimensional energy landscape for our beams.
For a three-mode truncation, the length constraint imposes the normalization condition: $a_0^2+a_1^2+a_2^2= 1$. 
This naturally motivates a parametrization using spherical coordinates $(\theta,\phi)$:
\begin{eqnarray}
a_0 &=& \cos{\theta}\cos{\phi}~,\nonumber \\
a_1 &=& \sin{\theta}\cos{\phi}~,\nonumber \\
a_2 &=&\sin{\phi}~\nonumber .
\end{eqnarray}
Within this coordinate system, we construct a two-dimensional energy landscape (Fig.~\ref{fig:2}b) that captures essential features of beam configurations independent of the compressive strain $\varepsilon$.

First, the left and right buckled states of the beam have mode amplitudes $(a_1,a_2,a_3)=(\pm 1,0,0)$,  which map to the stable fixed points at $(\theta,\phi)=(0,0)$ and $(\theta,\phi)=(\pi,0)$.
We stress that, due to the rescaling (Eq.~\ref{eq:rescale}), these fixed points are invariant under variations of $\varepsilon$. Second, the local energy minima are separated by an energy barrier that must be surmounted to induce a snap-through between the two buckled states; these correspond to saddle points that represent
purely antisymmetric 'S'-configurations\break   $((a_1,a_2,a_3)=(0,\pm1,0), ( \theta,\phi)=(\pm \pi/2,0))$ (Fig.~\ref{fig:2}b).  \break
Third, the presence of lateral pushers imposes geometric constraints that restrict the beam’s accessible configurations, giving rise to {\em excluded zones} and, as we will show later, additional stable equilibria.

\subsection{Evolution} 
\label{sec:evol}

We now show  how the landscape reveals the evolution of beam shapes under increasing compression or lateral motion of the pusher. We use a combination of gradient descent and sequential least squares programming methods to find the solve the model. First, the beam is in one of its free, buckled configurations. Then, as the pusher moves or $\varepsilon$ is increased, the excluded zones evolve accordingly and eventually
reach this configuration; the beam then makes contact with the pusher. 
Further increasing the driving leads to an overlap between the excluded zone and the free, buckled configuration. The system 
evolves, and the beam deforms due to its contact with the pusher. 
In such cases, the configuration lies either on the boundary of an excluded zone (single contact) or at the intersection of two such boundaries (dual contact).
As long as these configurations represent local minima in the allowed part of the energy landscape, the beam is stable, but upon further pushing or compressing, 
such minima become unstable and the beam configuration snaps to another stable configurations.

\begin{figure}[!t]
    \centering
    \includegraphics[]{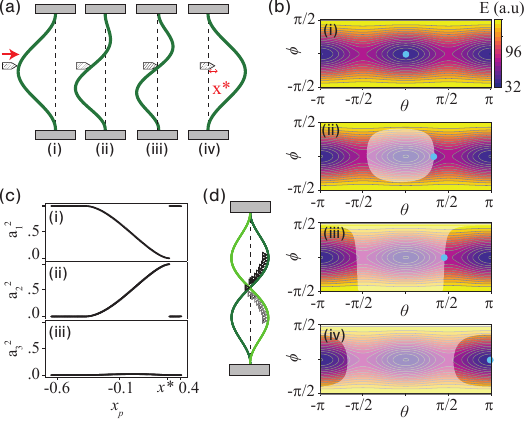}
    \caption{\textbf{Single-tip pusher.} 
    (a) Numerically computed beam configurations $x(z)$ and corresponding (b) energy landscapes for a beam under fixed $\varepsilon$, as a single-tip pusher moves laterally from left to right at fixed height $z_t=0.05$. (i-iv) Snapshots are shown for $x_p=-0.62\,$, $-0.21\,$, $-0.03\,$, and $0.14$.
    In the energy landscape, the cyan dot marks the current beam configuration and the white domain indicate the excluded zone imposed by the pusher. When the configuration is pushes passed the saddle point, the beam snaps to a new configuration (iv). 
    (c) Squared mode amplitudes as a function of pusher distance $x_p$ for $z_t=0.05$, showing that most energy is concentrated in the first two modes.
    (d) Beam shapes and corresponding pusher position at the onset of snap-through for a range of values of $z_t$, which shows that in each case, the configuration just before snapping is purely S-shaped (light green beam) or \MS-shaped (dark green beam), independent of $z_t$. The sign of $a_2$ is negative ($a_2 = -1$) if $z_t>0$ (black triangles, dark green beam), and positive ($a_2 = 1$) if $z_t<0$ (grey triangles, light green beam). 
    }\label{fig:2}
\end{figure}

We now illustrate this picture for 
a beam at fixed compression, laterally pushed by a single-tip pusher (Fig.~\ref{fig:2}a-b). 
We initialize the beam in the left buckled state  \break(Fig.~\ref{fig:2}ai,\ref{fig:2}bi). This configuration becomes inaccessible when the tip makes contact with the beam, creating an excluded zone that increases as $x_p$ increases (Fig.~\ref{fig:2}aii,\ref{fig:2}bii). Pushing further, the configuration reaches the saddle point corresponding to a mirrored S-shaped (or \MS-shaped) beam (Fig.~\ref{fig:2}aiii,\ref{fig:2}biii), after which the beam discontinuously \break
snaps and evolves to the oppositely buckled configuration (Fig.~\ref{fig:2}aiv,\ref{fig:2}biv) \cite{Chen21,Vangbo98,Pandey14}. 

Our simulations show that most of the elastic energy is stored in the first two modes of deformation: as the beam is laterally pushed ($x_p$ increases) the amplitude of the first mode decreases, while the amplitude of the second mode increases, the third mode remaining essentially at zero (Fig.~\ref{fig:2}c). When $a_2^2 = 1$, the configuration of the beam suddenly changes and $a_2$ jumps to zero, $a_1$ to $-1$: snap-through occurred and $x_p=x^*$ (Fig.~\ref{fig:2}c). Varying the $z_t$ position of the pusher does not qualitatively change this picture: with a single tip pusher the beam evolve along the $\phi = 0$ line, such that the unstable point lies at $(\theta=\pi/2, \phi=0)$. Hence, with a single tip pusher, snap-through occurs when the beam reaches either a S-shape or \MS-shape ($a_1=0$). This requires the pusher-tip to reach the centerline of beam (when $z_t=0$)\footnote{For the symmetric situation where $z_t=0$, whether the energy minimum ends up the S-shape or \MS-shape is set by spontaneous symmetry breaking, and the energy minimum remains at the line $\phi=0$; whereas for $z_t\ne0$ the shape is deterministically selected \cite{Chen21,Vangbo98,Pandey14}.}
or push beyond this centerline (when $z_t\neq0$; see Fig.~\ref{fig:2}d) \cite{Chen21,Vangbo98,Pandey14}. 
Hence, using a single tip pusher, snapping can be retarded, but not accelerated.

\section{Accelerated and two-step snapping with dual-tip pushers}\label{sec:results}

We now use our model to explore the beam evolution for dual-tip pushers. 
We consider three
different driving scenarios. 
In the first two, the vertical tip positions are fixed in the rescaled beam coordinates, which is theoretically simpler; we consider snapping induced by lateral pusher motion (scenario one), and by increased beam compression in the presence of a fixed pusher (scenario two). We also consider the experimentally
more realistic
third scenario, where the vertical tip positions are fixed in the lab-frame. 
We systematically vary the pusher parameters to obtain a detailed overview 
of the acceleration of the snapping, and 
where one-step and two-step snapping occur. 
We then perform the same analysis for scenario two and three (Sec.~\ref{sec:23}).
While differing in details, we find that
the overall mechanisms and 
even characteristic parameter ranges 
in
all scenarios are closely related.

\subsection{Three driving scenarios} \label{sec:geometries}
\begin{figure}[t]
\centering
\includegraphics{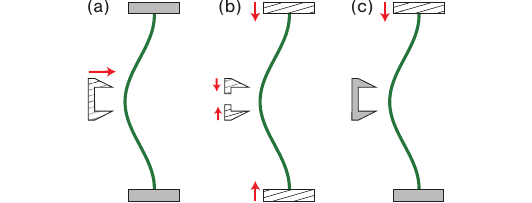}
    \caption{\textbf{Three scenarios.} (a) In scenario one, the beam is kept at constant compression $\varepsilon$ while the pusher moves laterally. (b) In scenario two,   
    the beam is compressed, the horizontal position of the pusher tips is fixed, and the vertical positions of the pusher tips are varied in the labframe so that their  
    rescaled heights $z_b$ and $z_t$ are fixed. (c) In scenario three, the beam is compressed while the horizontal and vertical positions of the pusher tips are fixed in the labframe.
    }\label{fig:3}
\end{figure}

We consider three driving scenarios in which we study snapping of a buckled beam
(Fig.~\ref{fig:3}). In the first scenario, the beam is kept under constant compressive strain $\varepsilon$, while the pusher moves horizontally towards the beam (Fig.~\ref{fig:3}a). In this case, $z_b$ and $z_t$ remain constant during driving, and the problem 
is symmetric under swapping $z_t\leftrightarrow-z_b$.
In the second and third scenario, the snapping is driven by increasing the axial compression of the beam, and the {\em horizontal} position of the pusher is fixed in the labframe, so that $x_p$ varies as
$X_p/(L\sqrt{\varepsilon})$ = 
$(X_p/L_0) \cdot 1/((1-\varepsilon)\sqrt{\varepsilon})$ 
(Eq.~\ref{eq:rescale}).

Furthermore, in the second scenario, we consider the theoretical simple case that 
$z_b$ and $z_t$ have fixed values. This 
implies that, in the labframe, 
the vertical pusher tip positions vary with the compression as $Z_{t,b}=(1-\varepsilon) L_0 z_{t,b}$. In this
scenario 
the swapping symmetry 
($z_t\leftrightarrow-z_b)$ is preserved
(Fig.~\ref{fig:3}b).
In the third scenario, we consider the experimental situation that the bottom plate and pusher positions are fixed in the labframe, so that  $Z_{t}$ and $Z_{p}$ are constants. Identifying the bottom and top vertical coordinate of the beam as $-L_0/2$ and $-L_0/2+L$ in the labframe, and -1/2 and 1/2 in the rescaled frame, we find that
$z_t=Z_t/L+\varepsilon/(2(1-\varepsilon))$ and $z_b=Z_b/L+\varepsilon/(2(1-\varepsilon))$. Hence, in this scenario the swapping symmetry is broken (Fig.~\ref{fig:3}c).

\subsection{Lateral moving pusher: Driving scenario one}
\label{sec:one}
We now discuss the snap-through behavior when the beam is under
constant compression and the pusher is moved laterally
(Fig.~\ref{fig:3}a).
When both tips are above (or below) $z=0$, only the tip closest to $z=0$ makes contact with the beam, yielding a behavior analogous to that of a single-tip pusher (Fig.~\ref{fig:2}). We therefore focus on the case where $z_t>0$ and $z_b<0$, so that both tips contact the beam. We consider a left-buckled beam, and denote the minimal horizontal location of a pusher, $x_p^*$, that allows for snapping. For appropriate tip positions, we observe accelerated snapping ($x_p^*<0$). Concomitant with this novel behavior, the beam shapes are more complex and require a third mode to accurately describe the deformation.

\subsubsection{One-step and two-step snapping}
\begin{figure}[t]
    \centering
    \includegraphics[]{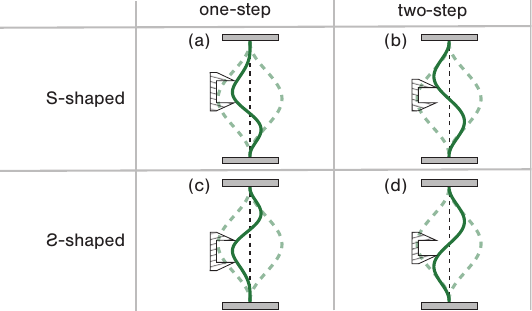}
    \caption{\textbf{Examples of the initial, final (dashed) and
    critical configuration just before snapping (full).}
    (a) For $(z_t,z_b)=(0.16,-0.03)$ the snapping occurs in one step at $x_p^*=-0.26$ -- both tips lose contact simultaneously. 
    In addition, as $z_t>|z_b|$, the top pusher tip contacts the beam first, forcing a S-shaped beam.
    (b) For $(z_t,z_b)=(0.10,-0.03)$ the snapping occurs in two steps, with the second step at $x_p^*=-0.22$ -- the tips lose contact sequentially. Similar to (a) the top pusher tip contacts the beam first yielding a S-shaped beam.
    (c-d) When we swap $z_t\leftrightarrow-z_b$, we observe \MS-shaped beams. 
    }\label{fig:4}
\end{figure}

As function of the locations of the pusher tips, two qualitatively distinct types of
snapping emerge. 
The snap-through consist of either {\em one step}, where both tips lose contact simultaneously (Fig.~\ref{fig:4}a), or {\em two steps}, where contact between tips and beam is lost sequentially (Fig.~\ref{fig:4}b).
Each scenario may occur with the beam adopting either an S-shaped configuration (with $a_2>0$) or a \MS-shaped configuration (with $a_2<0$; Fig.~\ref{fig:4}a-d).

The S and \MS~configurations are related by top-down symmetry, and are selected by the (broken) symmetry of the pusher, i.e., when
 $|z_b|\neq z_t$~\footnote{When $|z_b| = z_t$, the top-down symmetry is preserved, and the beam assumes a symmetric "W"-shape, snapping only after reaching a near-pure mode-three configuration \cite{Pandey14}.}.
In these asymmetric cases, the pusher tip closest to the center ($z=0$) first comes into contact with the beam and selects the sign of $a_2$;
for $|z_b|>z_t$, the beam adopts an S-shape with $a_2>0$ (Fig.~\ref{fig:4}a-b), while for
$|z_b|<z_t$ it assumes an \MS-shape with $a_2<0$ (Fig.~\ref{fig:4}c-d).

\begin{figure*}[t]
    \centering
    \includegraphics[]{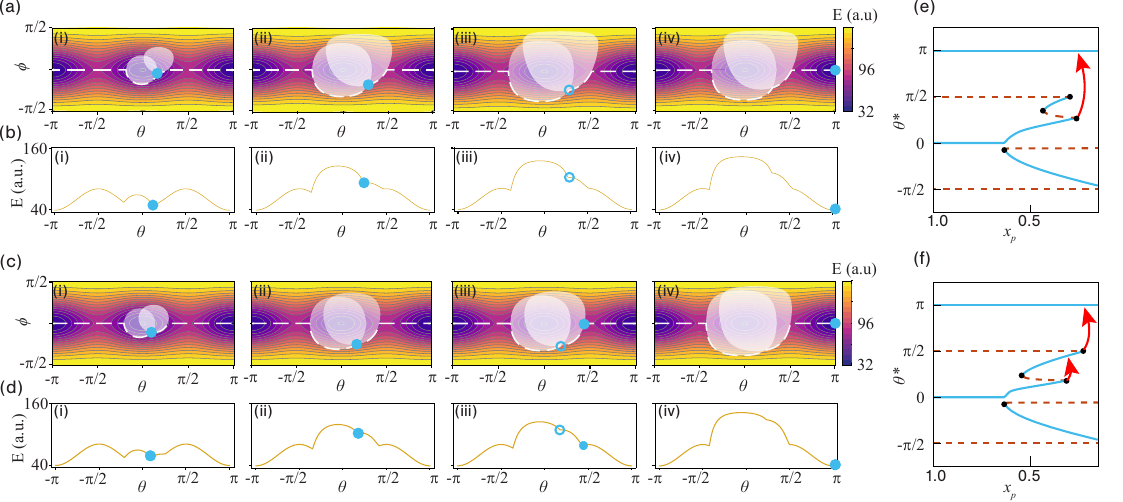}
    \caption{\textbf{Evolution of beam configuration for driving scenario one}. (a-b) {\em One-step snapping} for $(z_t, z_b)=(0.03,-0.16)$ with $x_p=[-0.55,-0.33,-0.28,-0.22]$ in (i-iv), respectively; here (a) shows the evolution of the configuration (blue) and excluded zones (grey areas), and (b) the energy along the curve of inherent minima. When the solution becomes unstable (iii), the nearby solution with $\phi=0$ is unstable, so that the beam loses contact with both pushers and evolves to the right-buckled configuration.
    (c-d) {\em Two-step snapping} for  $(z_t,z_b)=(0.03,-0.10)$ with $x_p=[-0.55,-0.36,-0.33,-0.22]$ in (i-iv), respectively. When the solution becomes unstable (iii), the nearby solution with $\phi=0$ is also stable, leading to the beam losing contact with only one tip, and a subsequent snapping at larger compression (iv).
    (e-f) Bifurcation diagrams illustrating the minima and maxima of the curve of inherent minima in blue and red dashed lines. Fold bifurcations are marked with black dots, and snapping events with red arrows.
    }\label{fig:5}
\end{figure*}

\subsubsection{One-step snapping: energy landscape interpretation}
We first analyze the emergence of the one-step scenario. We examine the energy landscape associated with the \MS-shaped beam\footnote{The evolution of an S-shaped beam follows by symmetry.} (Fig.~\ref{fig:4}c and  \ref{fig:5}a).
Starting from the left buckled state at $(\theta,\phi)=(0,0)$, the dual-tip pusher is laterally moved from left to right. Upon contact, the two tips create excluded zones in the energy landscape (Fig.~\ref{fig:5}ai). 

In the absence of the excluded zones,
all energy minima lie along the $\phi=0$ line ($\partial_\phi E(\theta, \phi) |_\theta = 0$ at $\phi=0$).
However, in the presence of the excluded zones, additional stable points can be found along the boundary, and their
stability is determined by the energy variation along these edges. 
Hence, all stable configurations reside along a curve that combines the relevant edges of the excluded zones and the 
$\phi=0$ line; 
we refer to this curve as the \textit{curve of inherent minima} (dashed line in Fig.~\ref{fig:5}ai,\ref{fig:5}bi).

When the beam makes contact with both tips of the pusher, the configuration becomes trapped in a local energy minimum located at the intersection of two excluded zones, which--unlike the single tip case--enables the beam to explore configurations beyond the $\phi=0$ line (Fig.~\ref{fig:5}ai).
As $x_p$ increases, the excluded zones expand and the beam initially remains trapped at their intersection (Fig.~\ref{fig:5}aii,\ref{fig:5}bii).
However, this minimum eventually becomes unstable as the gradient turns negative (open circles in Fig.~\ref{fig:5}aiii,\ref{fig:5}biii). This triggers the snap-through transition to the right-buckled configuration, where we note that 
in this case $x_p^*<0$, i.e. snapping occurs before the pusher reaches the centerline of the beam (Fig.~\ref{fig:5}aiv,\ref{fig:5}bvi).

\subsubsection{Two-step snapping: energy landscape interpretation}
In the two-step snapping scenario, the initial evolution is similar to the one-step case: the two tips trap the beam in a local energy minima situated at the intersection of the two excluded zones (Fig.~\ref{fig:5}ci-cii,\ref{fig:5}di-dii). The key difference is that 
when this local minima loses stability, another local minimum exists that traps the beam at the intersection of one excluded zone and the 
$\phi=0$ line (Fig.~\ref{fig:5}ciii,\ref{fig:5}diii). 
Here, the beam is still in contact with one of the pusher tips, and only by increasing $x_p$, the configuration reaches the saddle point at $(\theta,\phi)=(0,\pi/2)$  and the beams snaps (Fig.~\ref{fig:5}civ,\ref{fig:5}div).

Crucially, this second step is governed entirely by the pusher tip last in contact with the beam which is the tip located farthest from the middle (largest $|z|$). As a result, varying the position of the other pusher tip has no influence on the onset of this transition.
Despite being driven by a single pusher tip, this second transition occurs {\em before} the tip crosses the beam's centerline. This behavior is markedly different from the standard single-tip scenario and arises from the reversed shape of the beam and the corresponding reversed
sign of $a_2$. This reversed shape is caused by the tip that is closest to the center, and that, at this point of the evolution, is no longer in contact with the beam. Hence, although only the bottom tip is in contact with the beam at the onset of snapping, the beam is effectively constrained in the 'unnatural'
\MS-shaped configuration throughout this last stage of the evolution (see Fig.~\ref{fig:2}aii-aiii, Fig.~\ref{fig:4}d and Fig.~\ref{fig:5}c-d). 

\subsubsection{Bifurcation diagrams}
We can further represent 
the difference between one-step and two-steps snapping, by 
tracking the extrema along the curve of inherent minima as a function of $x_p$ for both scenarios (Fig.~\ref{fig:5}e-f).
The snapping in 
the one-step scenario and the first step of the two-step scenario correspond to the same instability. 
However, the key difference is that 
in the one-step scenario this initial transition evolves to the $\theta=\pi$
configuration, whereas in the
two-step scenario an intermediate state is reached (Fig.~\ref{fig:5}f). 
Together, these scenarios underscore that one-step and two-step snapping arise from the presence of multiple local energy extrema introduced by the second pusher tip. 

\subsubsection{Snapping threshold}
\begin{figure}[t]
    \centering
    \includegraphics[]{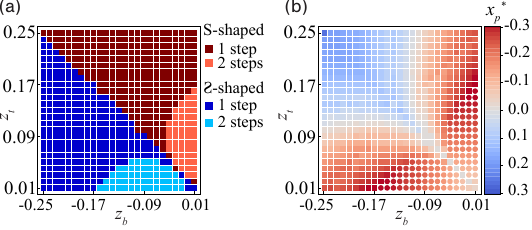}
    \caption{\textbf{Snapping in driving scenario one.}
    (a) State diagram for the four different scenarios as a function of ($z_t, z_b$). S-shaped (blue) and \MS-shaped (red) configurations obey a mirror symmetry along the line $z_t = -z_b$. One-step snapping is indicated in dark blue or red, two-step in light color.
    (b) Critical distance to snapping $x_p^*$ as a function of $(z_t,z_b)$
    (squares denote one-step snapping, circles two-step snapping).
    }\label{fig:6}
\end{figure}

Finally, we systematically investigate the influence of the tip heights, $z_t$ and $z_b$, on the snapping phenomonology and the critical snapping threshold to the right-buckled state $x_p^*$ 
(Fig.~\ref{fig:6}). 
As expected, the data obeys mirror symmetry along the line
$z_t=-z_b$, which delimits S-shaped and \MS-shaped scenarios (red and blue, respectively). We also observe that two-step snapping occurs within a small region of $z_t$, $z_b$ parameters, where both are small; for larger $z_t$, $z_b$, one-step snapping is observed .
The threshold $x^*_p$ is most negative near the boundary of these two regions and can reach values of the order of $-0.3$ (Fig.~\ref{fig:6}(b)).
Finally, while $x^*_p<0$ for a large range of $z$ parameters, pusher parameters with a large gap between their tips
(e.g., $z_t=0.2, z_b=-0.21$) may lead to single step snapping with $x_p^*>0$. Nevertheless, we note that $x^*_p$ for a dual-tip pusher is always less than for a single-tip pusher at either similar $z_t$ or similar $z_b$.
We conclude that accelerated snapping is a robust and tuneable feature when buckled beams are laterally pushed by a dual-tip pusher.

\subsection{Compression-driven snapping}\label{sec:23}
We now consider driving scenarios two and three, in which the lateral position of the pusher is fixed (i.e., $X_p$ is constant), and snapping is induced by increasing the axial compression $\varepsilon$ of the beam. 
In both scenarios, we consider lateral positions $X_p/L_0 = [-0.01,-0.03,-0.05]$ and quasistatically increase $\varepsilon$ until $\varepsilon < 0.12$ to remain in the small strain approximation. In both cases, accelerated snap-through can be triggered, following similar one-step and two-step responses.

\begin{figure*}[!t]
    \centering
    \includegraphics[]{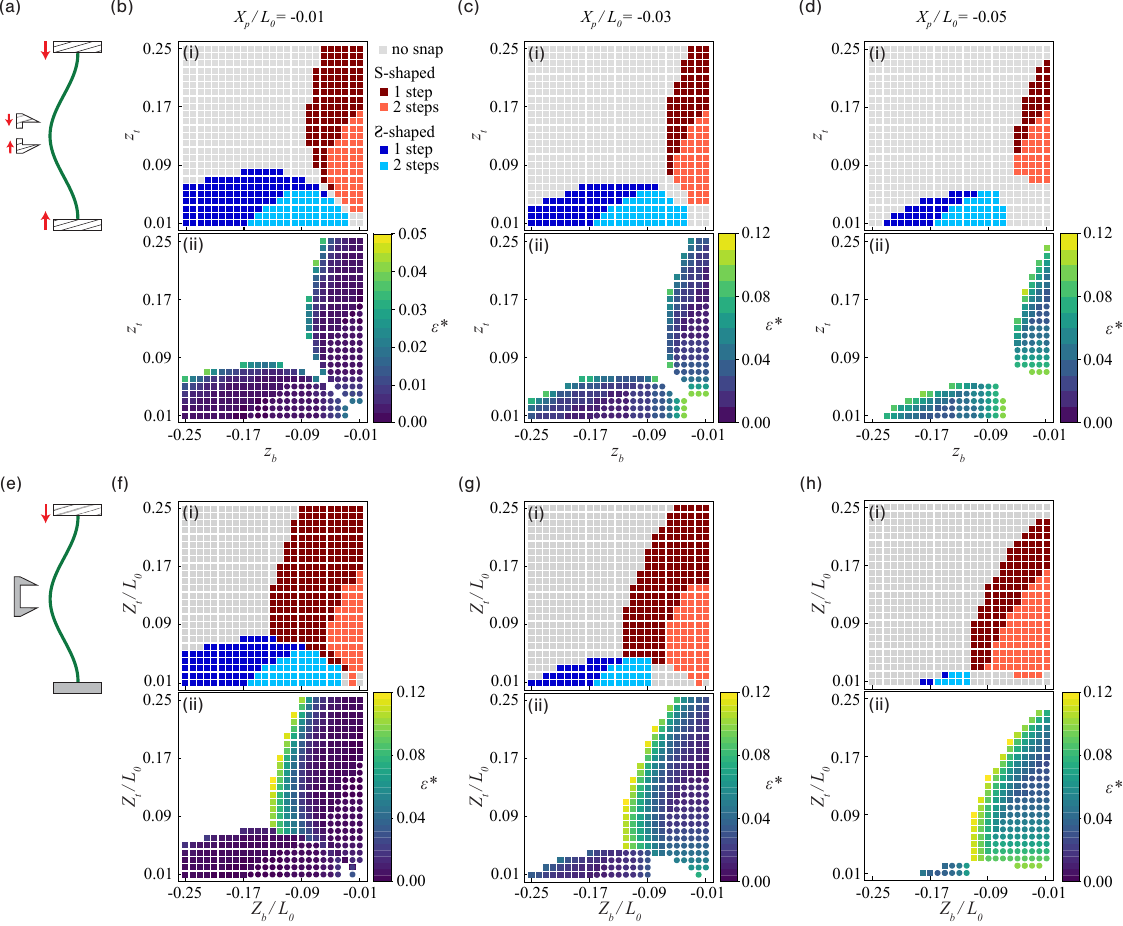}
    \caption{\textbf{Snap-through characterization for strain-driven beams.} 
    (a-d) Driving scenario two, in which the rescaled vertical positions of the pusher tips, $z_t$ and $z_b$, are constant throughout compression.
    (b-d) State diagrams (i) and critical strains $\varepsilon^*$ (ii), for 
    $X_P/L_0=-0.01$, $-0.03$, and $-0.05$, respectively. 
    (e-h) Driving scenario three, in which the dimensional tip heights are fixed in the labframe.
    (f-h) Corresponding state diagrams (i) and critical snap-through strains $\varepsilon^*$ (ii).
    }\label{fig:7}
\end{figure*}

\subsubsection{Driving scenario two}
We first consider the case in which we fix the rescaled vertical pusher positions, and systematically explore snapping for a range of $z_t$ and  $z_b$ (Fig.~\ref{fig:7}a-d). Both one-step and two-step snapping responses are observed, and the results
respect mirror (swap) symmetry
(Fig.~\ref{fig:7}b-d). In contrast to driving scenario one, a region without snapping emerges (gray zones in Fig.~\ref{fig:7});
when the fixed horizontal position is too far away from the centerline of the beam, no snapping can occur. This non-snapping region can be inferred from the data from scenario one (Fig.~\ref{fig:6}b); a necessary condition for snapping in scenario two is that 
$(|X_P|/L_0)<(1-\varepsilon)\sqrt{\varepsilon} |x_p^*|$, where
$x_p*$ is deduced in scenario one. 
Note that this expression implies that the farther the pusher is from the beam, the greater the compressive strain $\varepsilon$ required to trigger snap-through --- consistent with the trends observed in Fig.~\ref{fig:7}.

\subsubsection{Driving scenario three}
Finally, we examine the geometry where the pusher heights are rigidly connected to the stationary bottom plate.
We again find a wide range of configurations that trigger accelerated one-step and two-step snapping (Fig.~\ref{fig:7}f-h). As expected, the swap/mirror symmetry is broken and its
skewness increases with $X_p$. 
Consequently, compression often leads to snapping configurations where the bottom tip comes closer to the center of the beam than the top tip. 
Despite the added complexity of this geometry, the general trends of snapping are similar to those seen in scenario one and two.

\section{Conclusion}
We investigated the snapping behavior of slender buckled beams due to lateral forcing with two-tipped pushers.
Crucially, we showed that the introduction of a second pusher tip significantly enriches the deformation space, enabling both accelerated snapping and two-step snapping.
Our strategy opens a new route to 
advanced snapping, including in parameter regimes that were hitherto inaccessible, 
with applications in soft robots, smart sensors and \textit{in materia} computing \cite{togglerons}.

\section*{Acknowledgements}
We are grateful to M. Munck for early experiments in a related system and we acknowledge insightful discussions with L. Kwakernaak. We thank J. Mesman and D. Ursem for technical support. H.B., M.v.H and C.M.M. acknowledge funding from European Research Council Grant ERC-101019474. H.B. acknowledges funding from the European Union’s Horizon 2020 research and innovation programme under the Marie Sklodowska-Curie grant agreement number 101102728.

\bibliographystyle{elsarticle-num}
\bibliography{Snapping.bib}

\cleardoublepage
\appendix
\setcounter{figure}{0}

\section{Experimental mode decomposition}
To validate our assertion that the first three modes are dominant in describing the beam shape,
we perform a modal decomposition of the beam configuration in the representative experiments shown in Fig.~\ref{fig:1}.

Experiments are preformed with a beam of length $L_0=100\pm0.5\,$mm, in-plane thickness of $3\pm0.5\,$mm and out-of-plane width $20\pm0.5\,$mm. The beam is fabricated by casting two-component polyvinyl siloxane elastomer (Zhermack Elite double 22 with Young's modules $0.8\,$MPa and Poisson's ratio $\approx 0.5$) into a 3D-printed mold. After curing, the beam is removed and dusted with talc powder to minimize friction and prevent sticking. The lateral pushers are 3D printed and mounted on precision linear stages for accurate positioning and alignment.
Compression is applied using a custom-built apparatus designed for precise uniaxial loading, featuring high parallelism between the top and bottom plates, with a misalignment of less than \(6 \times 10^{-4}\) radians \cite{Kwakernaak2023}.  The axial compression is controlled by a stepper motor with an accuracy $\pm 0.01\,$mm and operated via an in-house LabView VI. 
Finally, the deformation of the beam is recorded using a CCD camera at $60\,$Hz with a resolution of $\approx 11\,$pixels/mm. The beam shape is extracted using open source software (ImageJ), and fitted to obtain the amplitudes $a_i$ of the first six modes:
\begin{equation}
    x_b(z_b) \approx \sum_{i=1}^{6}a_i\xi_i(a_b)~.
\end{equation}

\begin{figure}[t]
    \centering
    \includegraphics[]{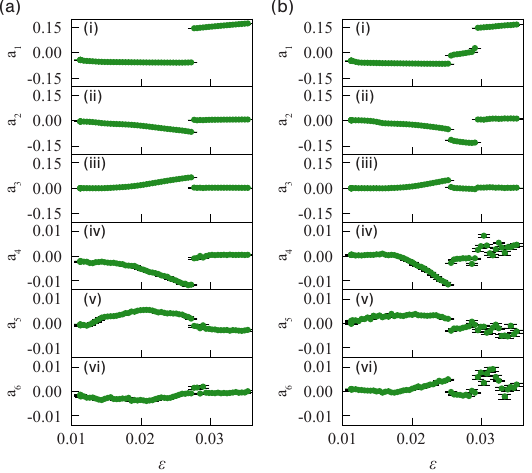}
    \caption{\textbf{Modal evolution of the beam.} 
    (a) Amplitude of the first six modes $a_1$-$a_6$  for the one-step scenario shown in Fig.~\ref{fig:1}b ($x_p^0=0.05$, $z_t^0=0.02$, and $z_b^0=-0.16$).
    Note the difference in vertical scale between panels (i-iii) and (iv-vi).
    (b) Similar, for the two-step scenario shown in Fig.~\ref{fig:1}c ($x_p^0=0.05$, $z_t^0=0.02$, and $z_b^0=-0.13$).
    }\label{fig:A1}
\end{figure}

We investigate both the one-step and two-step snapping behavior, by initializing the beam in the left-buckled state at $\varepsilon=0.011$ and quasistatically
(rate $10^{-3}\,$min$^{-1}$)
increasing the strain to $\varepsilon=0.035$. 
We find that the deformation energy is concentrated in the first three modes (Fig.~\ref{fig:A1}), and note that  after the first snapping event
in the two-step scenario the energy is concentrated in the first two modes (Fig.~\ref{fig:A1}b).
For other pusher geometries we have observed a similar dominance of the first three modes.
Hence, the experiments confirm the validity of truncating the modal expansion after the third mode.

\end{document}